\documentclass{elsart}

\usepackage{graphicx}
\usepackage{amsfonts}
\usepackage{latexsym}

\newcommand{\ave}[1]{\langle #1 \rangle}

\newcommand{\algname}[1]{{\sc #1}}
\newcommand{\keyw}[1]{{\bf #1}}
\newcommand{\comment}[1]{{$\rhd$ \it #1}}

\begin{document}

\begin{frontmatter}
  
  \title{A complete anytime algorithm for balanced number partitioning}
  \author{Stephan Mertens\thanksref{EMAIL}} \address{Institut f\"ur Theoretische
    Physik, Otto--von--Guericke--Universit\"at, 39106 Magdeburg, Germany}
  \thanks[EMAIL]{Email: stephan.mertens@physik.uni-magdeburg.de}

\begin{abstract}
  Given a set of numbers, the balanced partioning problem is to divide them into
  two subsets, so that the sum of the numbers in each subset are as nearly equal
  as possible, subject to the constraint that the cardinalities of the subsets
  be within one of each other.  We combine the balanced largest differencing
  method (BLDM) and Korf's complete Karmarkar-Karp algorithm to get a new
  algorithm that optimally solves the balanced partitioning problem.  For
  numbers with twelve significant digits or less, the algorithm can optimally
  solve balanced partioning problems of arbitrary size in practice. For numbers
  with greater precision, it first returns the BLDM solution, then continues to
  find better solutions as time allows.

\end{abstract}


\end{frontmatter}

\section{Introduction and overview}

The {\em number partitioning problem} is defined as follows: Given a list
$x_1,x_2,\ldots,x_n$ of non-negative, integer numbers, find a partition
$A\subset\{1,\ldots,n\}$ such that the {\em partition difference}
\begin{equation}
  \label{eq:cost-function}
  \Delta(A) = |\sum_{i\in A}x_i - \sum_{i\not\in A} x_i|,
\end{equation}
is minimized. In the constrained partition problem, the cardinality difference
between $A$ and its complement,
\begin{equation}
  \label{eq:magnetization}
  m = |A|-(n - |A|) = 2|A| - n,
\end{equation}
must obey certain constraints. The most common case is the {\em balanced
  partitioning problem} with the constraint $|m| \leq 1$.

Partitioning is of both theoretical and practical importance. It is one of Garey
and Johnson's six basic NP-complete problems that lie at the heart of the theory
of NP-completeness \cite{garey:johnson:79}.  Among the many practical
applications one finds multiprocessor scheduling and the minimization of VLSI
circuit size and delay \cite{coffman:lueker:91,tsai:92}.

Due to the NP-hardness of the partitioning problem \cite{karp:72}, it seems
unlikely that there is an efficient exact solution. Numerical investigations
have shown, however, that large instances of partitioning can be solved exactly
within reasonable time \cite{korf:95,gent:walsh:96,korf:98}. This surprising
fact is based on the existence of {\em perfect partitions}, partitions with
$E\leq 1$.  The moment an algorithm finds a perfect partition, it can stop.  For
identically, independently distributed (i.i.d.) random numbers $x_i$, the number
of perfect perfect partitions increases with $n$, but in a peculiar way. For $n$
smaller than a critical value $n_c$, there are no perfect partitions (with
probability one).  For $n > n_c$, the number of perfect partitions increases
exponentially with $n$.  The critical value $n_c$ depends on the number of bits
needed to encode the $x_i$.  For the unconstrained partitioning problem
\begin{equation}
  \label{eq:nc}
  n_c - \frac12\log_2 n_c = \frac12 \log_2\frac{\pi}{2}\ave{x^2},
\end{equation}
where $\ave{\cdot}$ denotes the average over the distribution of the $x_i$
\cite{mertens:98a}. The corresponding equation for the balanced partitioning
problem reads \cite{mertens:99b}
\begin{equation}
  \label{eq:ncb}
  n_c - \log_2 n_c = \log_2\left(\pi\sqrt{\ave{x^2}-\ave{x}^2}\right).
\end{equation}

For most practical applications the $x_i$ have a finite precision and
Eq.~\ref{eq:nc} resp.\ Eq.~\ref{eq:ncb} can be applied. Theoretical
investigations consider real-valued i.i.d.\ numbers $x_i\in[0,1)$, i.e.\ numbers with
infinite precision. In this case, there are no perfect partitions, and for a
large class of real valued input distributions, the optimum partition has a
median difference of $\Theta(\sqrt{n}/2^n)$ for the unconstrained resp.\ 
$\Theta(n/2^n)$ for the balanced case \cite{karmarkar:etal:86}. Using methods
from statistical physics, the average optimum difference has been calculated
recently \cite{mertens:98a,mertens:99b}. It reads
\begin{equation}
  \label{eq:Emin}
  \Delta_{\mathrm{opt}} = \sqrt{2\pi\ave{x^2}}\cdot\sqrt{n}\cdot 2^{-n}
\end{equation}
for the unconstrained and
\begin{equation}
  \label{eq:Eminb}
  \Delta_{\mathrm{opt}} = \pi\sqrt{\ave{x^2}-\ave{x}^2}\cdot n\cdot 2^{-n}
\end{equation}
for the balanced partioning problem. These equations also describe the case of
finite precision in the regime $1\ll n \ll n_c$.

For both variants of the partitioning problem, the best heuristic algorithms are
based on the Karmarkar-Karp differencing scheme and yield partitions with
expected $E=n^{-\Theta(\log n)}$ when run with i.i.d.\ real valued input values
\cite{karmarkar:karp:82,yakir:96}.  They run in polynomial time, but offer no
way of improving their solutions given more running time. Korf \cite{korf:98}
proposed an algorithm that yields the Karmarkar-Karp solution within polynomial
time and finds better solutions the longer it is allowed to run, until it
finally finds and proves the optimum solution.  Algorithms with this property
are referred to as {\em anytime algorithms} \cite{boddy:dean:89}.  Korf's
anytime algorithm is very efficient, especially for problems with moderate
values of $n_c$. For numbers $x_i$ with twelve significant digits or less ($n_c
\leq 33$), it can optimally solve partitioning problems of arbitrary size in
practice, since it quickly finds a perfect partition for $n>n_c$. For larger
values of $n_c$, several orders of magnitude improvement in solution quality
compared to the Karmarkar-Karp heuristic can be obtained in short time.

For practical applications of this NP-hard problem, this is almost more than one
might expect. Korf's algorithm is not very useful to find the optimum
constrained partition, however.  In this paper, we describe a modification of
Korf's algorithm, which is as efficient as the original, but solves the
constrained partition problem.  The next section comprises a description of
Korf's algorithm and the modifications for the balanced problem.  In the third
section we discuss some experimental results. The paper ends with a summary and
some conclusions.

\section{Algorithms}

\subsection{Differencing heuristics}

The key ingredient to the most powerful partition heuristics is the differencing
operation \cite{karmarkar:karp:82}: select two elements $x_i$ and $x_j$ and
replace them by the element $|x_i-x_j|$. Replacing $x_i$ and $x_j$ by
$|x_1-x_2|$ is equivalent to making the decision that they will go into opposite
subsets. Applying differencing operations $n-1$ times produces in effect a
partition of the list $x_1,\ldots,x_n$. The value of its partition difference is
equal to the single element left in the list.

Various partitions can be obtained by choosing different methods for selecting
the pairs of elements to operate on. In the {\em paired differencing method}
(PDM), the elements are ordered.  The first $\lfloor n/2\rfloor$ operations are
performed on the largest two elements, the third and the fourth largest, etc..
After these operations, the left-over $\lceil n/2 \rceil$ elements are ordered
and the procedure is iterated until there is only one element left.

Another example is the {\em largest differencing method} (LDM). Again the
elements are ordered.  The largest two elements are picked for differencing. The
resulting set is ordered and the algorithm is iterated until there is only one
element left.

For $1 \ll n \ll n_c$, i.e.\ in the regime where there are no perfect
partitions, and for random i.i.d.\ input numbers, the expected partition
differences are $\Theta(n^{-1})$ for PDM \cite{lueker:87} and $n^{-\Theta(\log
  n)}$ for LDM \cite{yakir:96}.

LDM, being superior to PDM, is not applicable to the constrained partioning
problem. PDM on the other hand yields only perfectly balanced partitions. Yakir
proposed a combination of both algorithms, which finds perfectly balanced
partitions, but with an expected partition difference of $n^{-\Theta(\log n)}$
\cite{yakir:96}. In his {\em balanced LDM} (BLDM), the first iteration of PDM is
applied to reduce the original $n$-element list to $\lceil n/2 \rceil$ elements.
By doing so, it is assured that the final partition is balanced, regardless of
which differencing operations are used thereafter. If one continues with LDM, a
final difference of $n^{-\Theta(\log n)}$ can be expected.

The time complexity of LDM, PDM and BLDM is $O(n\log n)$, the space-complexity
is $O(n)$.

\subsection{Korf's complete anytime algorithm}

\begin{figure}[htbp]
  \includegraphics[width=\columnwidth]{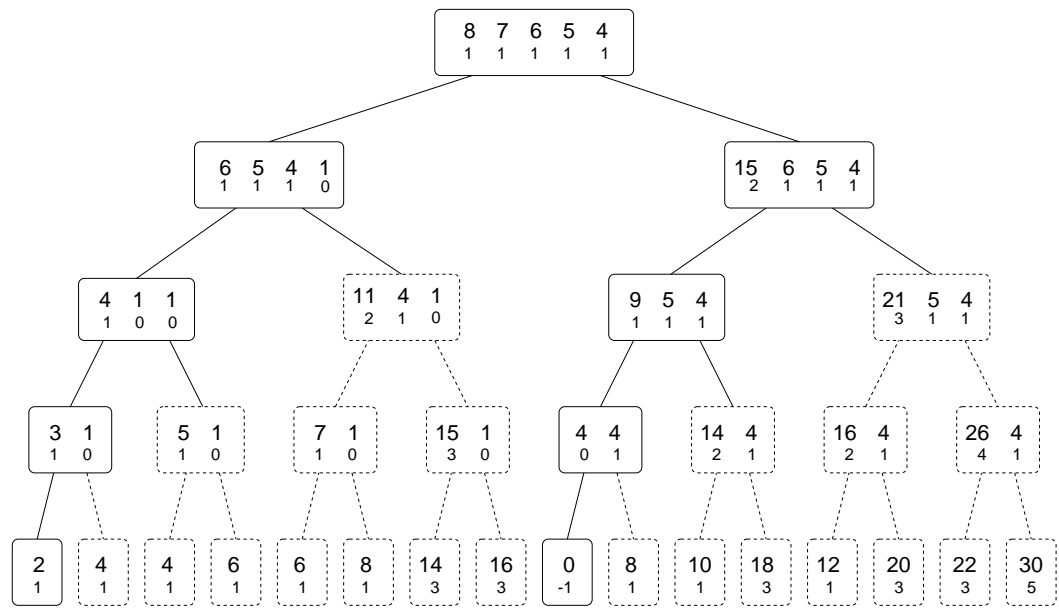}
  \caption{Tree generated by complete Karmarkar-Karp differencing on the list $8,7,6,5,4$.
    Left branch: Replace the two largest numbers by their difference.  Right
    branch: Replace the two largest numbers by their sum.  The numbers in small
    font are the effective cardinalities needed to keep track of the cardinality
    difference of the final partition.  The dashed parts of the tree are pruned
    by the algorithm.}
  \label{fig:example}
\end{figure}

LDM and BLDM are the best known heuristics for the partioning problem, but they
find approximate solutions only.  Korf \cite{korf:98} showed, how the LDM can be
extended to a {\em complete anytime algorithm}, i.e.\ an algorithm that finds
better and better solutions the longer it is allowed to run, until it finally
finds and proves the optimum solution: At each iteration, the LDM heuristic
commits to placing the two largest numbers in different subsets, by replacing
them with their difference.  The only other option is to place them in the same
subset, replacing them by their sum. This results in a binary tree, where each
node replaces the two largest remaining numbers, $x_1\geq x_2$: the left branch
replaces them by their difference, while the right branch replaces them by their
sum:
\begin{equation}
  x_1, x_2, x_3, \ldots \mapsto \left\{
  \begin{array}{rl}
    |x_1-x_2|, x_3, \ldots & \mbox{ left branch } \\
    x_1+x_2, x_3, \ldots & \mbox{ right branch } 
  \end{array}
  \right.
\end{equation}
Iterating both operations $n-1$ times generates a tree with $2^{n-1}$ terminal
nodes.  The terminal nodes are single element lists, whose elements are the
valid partition differences $\Delta$.  Korf's {\em complete Karmarkar-Karp}
(CKK) algorithm searches this tree depth-first and from left to right. CKK first
returns the LDM solution, then continues to find better solutions as time
allows.  See Fig.~\ref{fig:example} for the example of a tree generated by CKK.

There are two ways to prune the tree: At any node, where the difference between
the largest element in the list and the sum of all other elements is larger than
the current minimum partition difference, the node's offspring can be ignored.
If one reaches a terminal node with a perfect partition, $\Delta \leq 1$, the
entire search can be terminated. The dashed nodes in Fig.~\ref{fig:example} are
pruned by these rules.

In the regime $n < n_c$, the number of nodes generated by CKK to find the
optimum partition grows exponentially with $n$. The first solution found, the
LDM-solution, is significantly improved with much less nodes generated, however.
In the regime $n > n_c$, the running time decreases with increasing $n$, due to
the increasing number of perfect partitions. For $n\gg n_c$, the running time is
dominated by the $O(n\log n)$ time to construct the LDM-solution, which in this
regime is almost always perfect.

\subsection{A complete anytime algorithm for constrained partioning}

The application of differencing and its opposite operation leads to lists, in which
single elements represent several elements of the original list.  In order to
apply CKK to the constrained partitioning problem, one needs to keep track of
the resulting cardinality difference. This can be achieved by introducing an
{\em effective cardinality} $m_i$ for every list element $x_i$. In the original
list, all $m_i=1$. The differencing operation and its opposite become
\begin{equation}
  \label{eq:branching-2}
  {{x_1} \atop {m_1}}, {{x_2} \atop {m_2}}, {{x_3} \atop {m_3}}, \ldots \mapsto
  \left\{
  \begin{array}{rl}
    {{|x_1-x_2|} \atop{m_1-m_2}}, {{x_3} \atop {m_3}}, \ldots & \mbox{ left branch } \\
    {{x_1+x_2}\atop{m_1+m_2}},  {{x_3} \atop {m_3}}, \ldots & \mbox{ right branch}
  \end{array}
  \right..
\end{equation}
Fig.~\ref{fig:example} shows how the $m_i$ evolve if the branching rule
\ref{eq:branching-2} is applied to the list $8,7,6,5,4$. The terminal nodes
contain the partition difference and the cardinality difference. A simple
approach to the constrained partition problem is to apply CKK with the branching
rule \ref{eq:branching-2} and to consider only solutions with matching $m$. This
can be very inefficient, as can be seen for the constraint $m=n$. This extreme
case is trivial, but CKK needs to search the complete tree to find the solution!

As a first improvement we note, that an additional pruning rule can be applied.
Let $m_{\mathrm{max}} := \max_i\{|m_i|\}$ and $M := \sum_i |m_i|$ at a given
node. The cardinality difference $m$ which can be found within the offspring of
this node, is bounded by
\begin{equation}
  \label{eq:m_bound}
  2 m_{\mathrm{max}} - M \leq |m| \leq M.
\end{equation}
Comparing these bounds to the cardinality constraint, one can prune parts of the
tree. Consider again the case $m=n$ as an example: The trivial solution is now
found right away.

CKK finds the first valid partition (the LDM solution) after generating $n$
nodes.  For the constrained partition problem, this can not be guaranteed --
except in the case of balanced partitions, where we can use the BLDM strategy.
Applying the first $\lfloor n/2 \rfloor$ PDM operations to the original list
leaves us with a $\lceil n/2 \rceil$-element list with all $m_i=0$ (resp.~with a
single $m_i=1$ if $n$ is odd). CKK applied to this list produces only perfectly
balanced partitions, the BLDM solution in first place. To keep the completeness
of the algorithm, we have to consider the alternative to each of the PDM
operations, i.e.\ to put a pair of subsequent numbers in the same subset.

\begin{figure}[htbp]
  \fbox{ \parbox[t]{\columnwidth}{
\begin{tabbing}
  \quad \=\quad \=\quad \=\quad \=\quad \=\quad \=\quad \=\quad \=\quad \=\quad \=\quad \kill
  \algname {Complete-BLDM} $\Big((x_1,m_1),(x_2,m_2),\ldots,(x_k,m_k)\Big)$ \\
  \>\comment{Called with sorted list $x_1\geq x_2\geq \ldots \geq x_n$ and all $m_i=1$}\\
  \>\comment{the minimum partition difference $\Delta$ among all partitions with}\\
  \>\comment{cardinality-difference $m$ is returned.}\\
  \>\keyw{if} $k = n$ \keyw{then} \comment{initialize} \\
  \> \> $\Delta := \infty$; \\
  \> \keyw{fi} \\
  \>\keyw{if} $k = 1$ \keyw{then} \comment{terminal node} \\
  \> \> \keyw{if} $|m_1| = |m|$ \keyw{and} $x_1 < \Delta$ \keyw{then}  \\ 
  \> \> \> $\Delta := x_1$; \comment{found a better solution}\\
  \> \> \keyw{fi} \\
  \>\keyw{else} \\
  \> \> \comment{pruning based on partition- and cardinality difference}\\
  \> \> \keyw{if} $2\cdot\max_i\{x_i\}  - \sum_i x_i \geq \Delta$ \keyw{return}; \\
  \> \> \keyw{if} $2\cdot\max_i\{|m_i|\} - \sum_i|m_i| > |m|$ \keyw{or} $\sum_i|m_i| < |m|$
        \keyw{return}; \\
  \> \> \keyw{if} $k <= \lceil n/2 \rceil$ \keyw{then} \comment{LDM phase}\\
  \> \> \> \comment{sort list such that $x_1 \geq x_2 \geq \cdots \geq x_k$} \\
  \> \> \> \algname{Sort}$\Big((x_1,m_1),(x_2,m_2),\ldots,(x_k,m_k)\Big)$; \\
  \> \> \keyw{fi} \\
  \> \> \comment{branch}\\
  \> \> \algname{Complete-BLDM} $\Big((x_3,m_3), \ldots (x_k,m_k), (x_1-x_2, m_1-m_2)\Big)$; \\
  \> \> \algname{Complete-BLDM} $\Big((x_3,m_3), \ldots (x_k,m_k), (x_1+x_2, m_1+m_2)\Big)$; \\
  \> \keyw{fi}
\end{tabbing}
}}
  \caption{Complete BLDM algorithm to solve the constrained partition problem.}
  \label{fig:alg}
\end{figure}

An outline of the complete BLDM algorithm can be seen in Fig.~\ref{fig:alg}.
Note that in an actual implementation several modifications should be applied to
improve the performance. Instead of sorting the list at every node in the LDM
phase, it is much more efficient to sort only when switching from PDM to LDM and
insert the new element $x_1\pm x_2$ in the LDM-phase such that the order is
preserved. The $\max$ and $\sum$ of $x_i$ and $|m_i|$ should be calculated only
once and then locally updated when the list is modified

\section{Experimental results}

We implemented the complete BLDM algorithm to test its performance as an exact
solver, a polynomial heuristic and an anytime algorithm. For all computer
experiments we use i.i.d.\ random numbers $x_i$, uniformly distributed from $0$
to $2^b-1$, i.e.\ $b$-bit integers.

\begin{figure}[htbp]
  \includegraphics[width=\columnwidth]{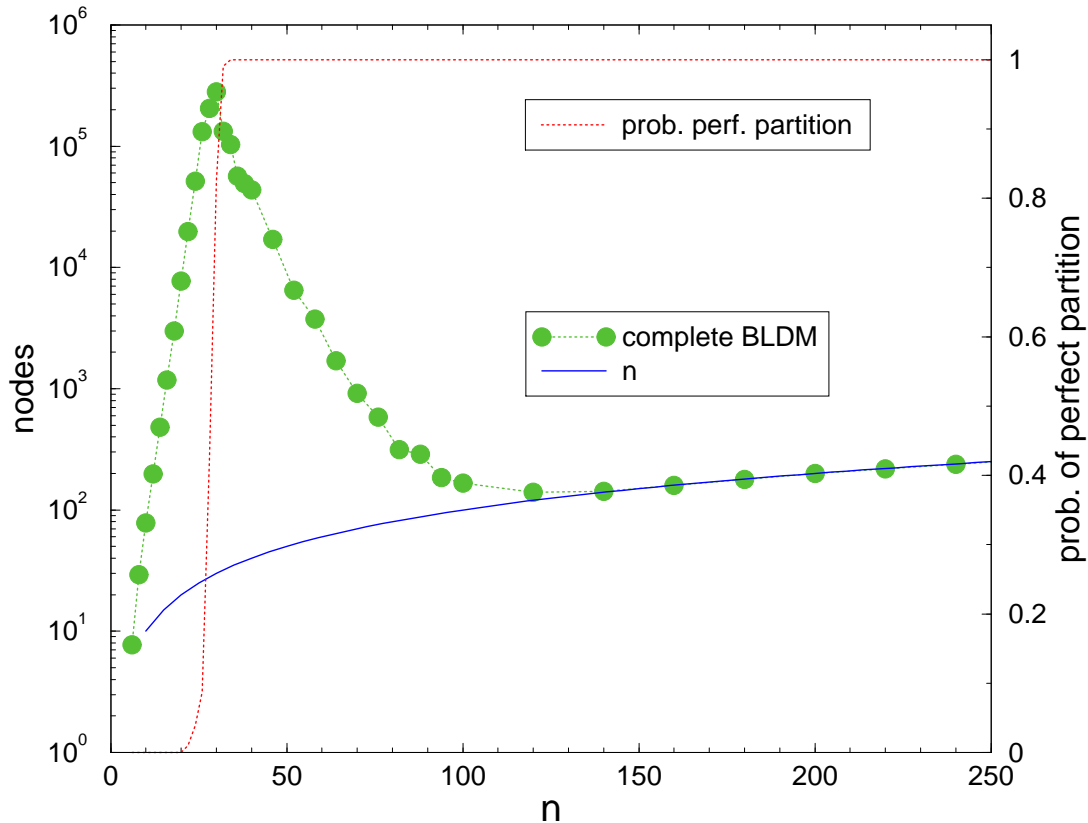}
  \caption{Number of nodes generated by the complete BLDM algorithm to optimally 
    partition random 25-bit integers.}
  \label{fig:korf}
\end{figure}

To measure the performance of the algorithm as an exact solver, we count the
number of nodes generated until the optimum solution has been found and proven.
The result for $25$-bit integers is shown in Fig.~\ref{fig:korf}.  Each data
point is the average of $100$ random problem instances. The horizontal axes
shows the number of integers partitioned, the vertical axes show the number of
nodes generated (left) and the fraction of instances that have a perfect
partition (right).  Note that we counted all nodes of the tree, not just the
terminal nodes.  We observe three distinct regimes: for $n < 30$, the number of
nodes grows exponentially with $n$, for $n > 30$ it decreases with increasing
$n$, reaching a minimum and starting to increase again slowly for very large
values of $n$.

Eq.~\ref{eq:ncb} yields $n_c=29.7$ for our experimental setup, in good agreement
with the numerical result, that the probability of having a perfect partition is
one for $n \geq 30$ and drops sharply to zero for smaller values of $n$. In the
regime $n < n_c$, the algorithm has to search an exponential number of nodes in
order to prove the optimality of a partition. For $n > n_c$ it finds a perfect
partition and stops the search prematurely. The number of perfect partitions
increases with increasing $n$, making it easier to find one of them. This
explains the decrease of searching costs.  For $n \gg n_c$, the very first
partition found already is perfect. The construction of this BLDM solution
requires $n$ nodes.

\begin{figure}[htbp]
  \includegraphics[width=\columnwidth]{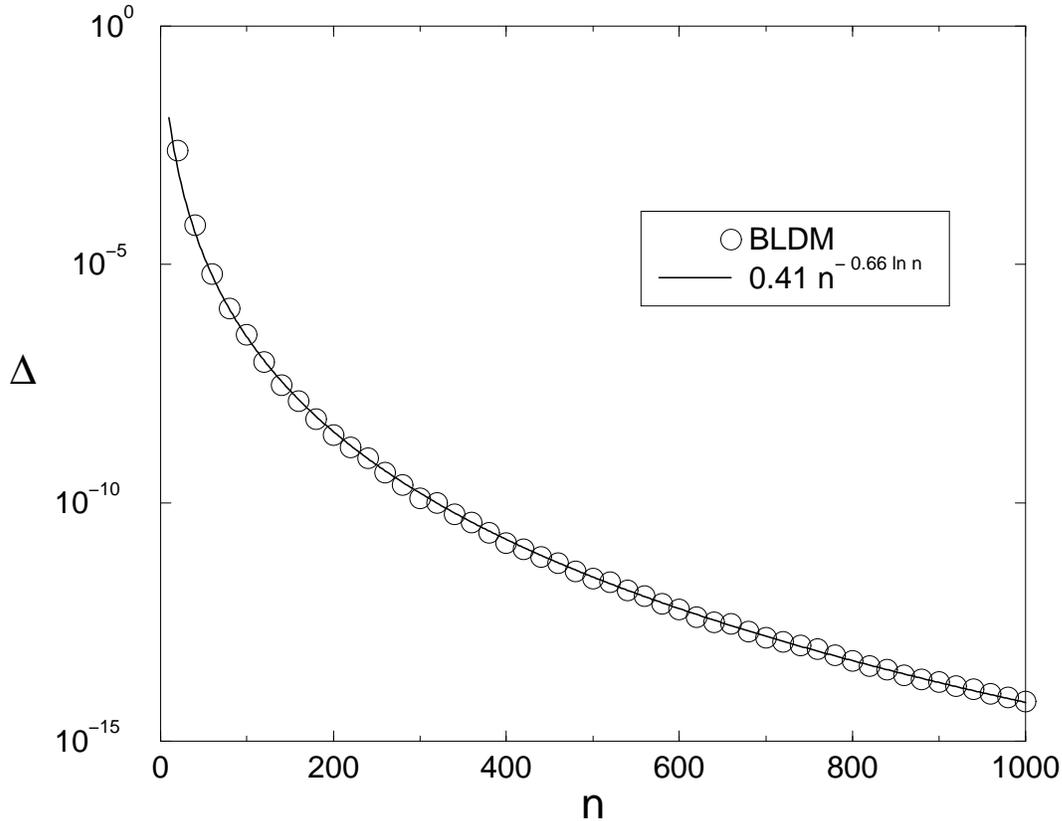}
  \caption{Partition difference found by heuristic BLDM for ``infinite precision numbers'' from
    the interval $[0,1)$.}
  \label{fig:ckk-heuristic}
\end{figure}

We have seen that for $n \gg n_c$ the BLDM heuristic yields perfect partitions.
How does it behave in the other extreme, the ``infinite precision limit'', $n
\ll n_c$?  Yakir \cite{yakir:96} proved that in this limit BLDM yields an
expected partition difference of $n^{-\Theta(\log n)}$. For a numerical check we
applied BLDM to partition $2n$-bit integers to ensure that $n \ll n_c$. The
partition difference is then divided by $2^{2n}$ to simulate infinite precision
real numbers from the interval $[0,1)$.  Fig.~\ref{fig:ckk-heuristic} shows the
resulting partition difference.  Each data point is averaged over $1000$ random
instances. Due to the numerical fit in Fig.~\ref{fig:ckk-heuristic} it is
tempting to conjecture
\begin{equation}
  \label{eq:BLDM-conjecture}
  \Delta_{\mathrm{BLDM}} = (\sqrt{2}-1) n^{-\frac23\ln n}.
\end{equation}

\begin{figure}[htbp]
  \includegraphics[width=\columnwidth]{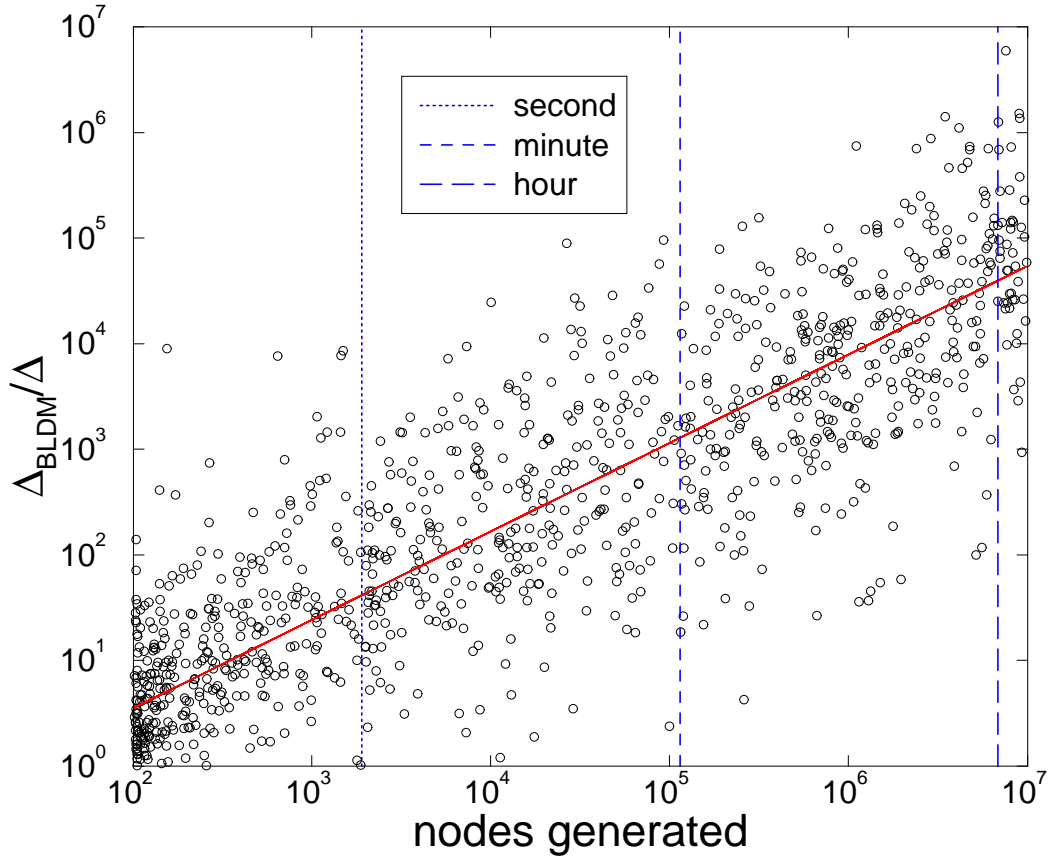}
  \caption{Solution quality relative to BLDM solution for 100 random 150-bit integers. Data points are
    shown for runs on 100 random instances. The solid line is a numerical fit.}
  \label{fig:progress}
\end{figure}

If we want better solutions than the BLDM solution we let the complete BLDM run
as long as time allows and take the best solution found. We applied this
approach to partition $100$ random $150$-bit integers.  Perfect partitions do
not exist (with probability one), and the true optimum is definitely out of
reach.  The results can be seen in Fig.~\ref{fig:progress}. The horizontal axis
is the number of nodes generated, and the vertical axes is the ratio of the
initial BLDM solution to the best solution found in the given number of node
generations, both on a logarithmic scale. The entire horizontal scale represents
about 90 minutes of real time, measured on a Sun SPARC 20.  The fact that the
number of nodes per second is a factor of $1000$ smaller than reported by Korf
for the CKK \cite{korf:98} on $48$-bit integers is probably due to the fact that
we had to use a multi-precision package for the $150$-bit arithmetic while Korf could
stick to the fast arithmetic of built-in data types. Even with this slow node
generation speed we observe a several order of magnitude improvement relative to
the BLDM solution in a few minutes.  A least square fit to the data of $100$
runs gives
\begin{equation}
  \label{eq:least-square}
  \frac{\Delta_{\mathrm{BLDM}}}{\Delta} \approx 0.075 (\mathrm{\#nodes})^{0.84},
\end{equation}
but the actual data vary considerably.

\section{Summary and conclusions}

The main contribution of this paper is to develop a complete anytime algorithm
for the constrained number partioning problem. The complete Karmarkar-Karp
algorithm CKK, proposed by Korf for the unconstrained partitioning problem, can
be adapted to the constrained case simply by keeping book of the effective
cardinalities and by extending the BLDM heuristic to a complete algorithm. The
first solution the algorithm finds is the BLDM heuristic solution, and as it
continues to run it finds better and better solutions, until it eventually finds
and verifies an optimal solution.

The basic operation of the complete BLDM is very similar to Korf's CKK. The
additional processing of the effective cardinalities has only a minor impact on
the runtime. The pruning based on estimating the cardinality difference leads to
a gain in speed, on the other hand.  Therefore we adopt Korf's claim: For
numbers with twelve significant digits or less, complete BLDM can optimally
solve balanced partitioning problems of arbitrary size in practice.

\bibliographystyle{plain}

\bibliography{complexity,cs}

\end{document}